\documentstyle[aps,preprint,psfig]{revtex}
\renewcommand{\psdraft}{}
\begin{document}

\title{$J/\psi$ AND $\psi'$ SUPPRESSION IN HADRONIC MATTER}

\author{K. Martins}
\address{MPG Arbeitsgruppe  ''Theoretische Vielteilchenphysik''\\
         Universit\"at Rostock, D-18051 Rostock, Germany}

\maketitle

\vspace{1cm}

\noindent MPG--VT--UR 77/95\\
\noindent {\sc December} 1995\\
\noindent to be published in Prog. Part. Nucl.Phys.
\begin{abstract}
We present a microscopic calculation of the breakup cross sections of $J/\psi$
 and $\psi'$ on pions and nucleons as a function of the kinetic energy.
These cross sections are used for the investigation of the $J/\psi$ to
continuum and $\psi'/J/\psi$ ratios in ultrarelativistic heavy ion collisions.
The contribution of produced comoving pions to the $\psi'/J/\psi$ signal is
calculated.
While this model can account for the data, the uncertainties in the parameter
values do not allow to exclude
 the possibility of additional sources for charmonium absorption, like
a resonance gas or the quark gluon plasma.

\vspace{1em}

\centerline{\large Keywords}
\parindent0cm

heavy-ion collisions; quark potential model; hadron-hadron interaction; 
$J/\psi$ suppression
\end{abstract}

%
\newpage
Charmonium states are possible probes for dense matter formation in the early 
phase of ultrarelativistic heavy ion collisions.
The measurement of the $J/\psi$ signal was proposed as a test for deconfined
matter (Matsui and Satz, 1986). 
However, it has been shown that the $J/\psi$ yield is affected by many 
processes, which obscure the data analysis.
The $\psi'$ to  $J/\psi$ ratio seems to give a much clearer signal, since 
the production process is the same for both particles.

We consider the absorption of $J/\psi$ and $\psi'$ mesons by dissociation on
hadrons in a microscopic approach. The quark potential model is used
for the description of hadron--hadron interactions, for more details see
(Barnes and Swanson 1992, Martins {\it et al.} 1995).
Mesons are described as quark-antiquark bound states. The interaction 
Hamiltonian $H^{I}$
between a quark and antiquark is given by
a Coulomb--like term and the spin-spin  interaction that arise
from one--gluon exchange. An additional term $H^{np}$ accounts for the
nonperturbative nature of strong interaction at low energies.
The parameters of our model are fitted
to the mass spectrum of the relevant hadrons ($\pi, \rho , D , D^* , J/\psi, 
\psi', p, \Lambda_c, \Sigma_c$).
The charmonium dissociation can be understood as a quark rearrangement process
of the form $
  (Q \bar Q) + (q \bar q) \to (Q \bar q) + ( q \bar Q)$
   for dissociation on light mesons and
  $(Q \bar Q) + (q d) \to (Q d) + ( q \bar Q)$ for dissociation on
   nucleons
where $Q, q$ represent a heavy and light quark respectively and $d$ a diquark.
We calculate the transition matrix element from initial state 
(hadrons $A$ and $B$) to the final state  (hadrons $C$ and $D$) as the
 sum over all possible quark exchange subprocesses between
the hadrons, restricting ourselves to Born approximation.
The resulting cross sections are shown in Fig.\,1, where only
the final channels with lowest thresholds are included.
The reaction probability is enhanced near the reaction threshold.
In addition to the threshold behaviour, the asymptotic cross sections at
very high energies can be calculated from only the perturbative part of $H^I$.
In the intermediate energy range, 
the behaviour is  interpolated between both limiting cases.
The $\psi'$ has a lower reaction threshold and a higher peak value than the 
$J/\psi$ and therefore it is stronger absorbed in hadronic matter. 

$pA$ data (Alde {\it et al.}, 1991) show a $\psi'/J/\psi$ ratio which is
nearly independent of the mass number $A$. A charmonium nucleus absorption 
cross section of 4.5 mb for both $J/\psi$ and $\psi'$ is capable to reproduce 
the data.
Due to finite formation times for the $\psi$ states, they are
formed far outside the nucleus. 
The color singlet absorption of such a small expanding  
$Q \bar Q$ state is not sufficiently
large to account for the nucleonic absorption. One has to conclude
that there is another absorption mechanism, which might be the color octet 
interaction of the $Q \bar Q$ with nucleons.

A quantitative investigation of the measured suppression ratio 
(Ramos, 1995) has been done recently 
 in a  the row on row model of the nucleus--nucleus scattering
(Wong, 1995).
It has been shown, that the same $\psi'/\psi$ ratio in S-U collisions can
be obtained with an absorption on soft gluons, on produced comoving hadrons
or in a mixed scenario.
The cross section at a given impact parameter $b$ is given by
\vspace{-0.5em}
\begin{eqnarray}
  \label{sigmaB}
  {d \sigma^{AB}_{\psi}({\bf b})} \over {\sigma^{NN}_{\psi} d {\bf b}}
 & = & \int {d {\bf b}_A \over (\sigma_{abs}^{\psi N})^2 }
  \left\{ 1- \left[  1- T_A({\bf b}_A)\sigma_{abs}^{\psi N}\right]^A\right\}
    \left\{ 1- \left[ 1- T_B({\bf b}\mbox{-}{\bf b}_A)
     \sigma_{abs}^{\psi N}\right]^B \right\}
F_{co}({\bf b}, {\bf b}_A)
\\
     F_{co} ({\bf b}, {\bf b}_A) 
 & = & {1\over N_< N_>} \sum_{n=1}^{N_<} a_n \sum_{i=1}^n 
       \exp\left[\int_{t_0(n)}^{t_f}d\, t {1 \over \tau_\psi(t)}\right].
\end{eqnarray}

$N_<$ and $N_>$ are the smaller and larger number of nucleons in the 
colliding rows at transversal location ${\bf b} + {\bf b}_A$, $T_A$ and 
$T_B$ are the thickness functions of the nuclei and n is the number of volume 
elements with $n$ collisions inside. The comover absorption $F_{co}$ is mainly 
determined by the $\psi$ relaxation time $t_\psi$.
An analog equation holds for the $\psi'$ cross section.
Assuming a constant life time, we calculate $t_\psi$
from the absorption cross sections with a surrounding of
thermal pions at 200 and 250 MeV temperature.

The results for the contribution of secondary pions to the $J/\psi$ absorption
in A-B collisions provide a good description of the data.
Using the same parameters, we perform the same calculation for 
the $\psi' / \psi$ ratio. 
This ratio is more
sensitive to the interaction of the quarkonium state with comoving matter
since the influence of nucleons almost cancels in this ratio.

Our result is shown in Fig.\,2. At lower temperatures the
$\psi'/\psi$ ratio becomes larger. For a reasonable temperature of 200 MeV, 
there is room for an additional absorption mechanism. 
However, this are not  necessarily hints
for an evidence of a quark gluon plasma,  $\rho$ resonances and other
correlated matter states are also capable to dissolve the charmonium states 
because of the low reaction threshold for these processes.
Some more detailed investigation has to be done in order to understand the
true nature of charmonium suppression before one can conclude
the evidence of a quark gluon plasma phase.
\vspace{1.5em}

\centerline{REFERENCES}

\leftmargin4cm
\noindent
Matsui, T. and Satz, H. (1986).
      {\it Phys. Lett.} {\underline {B 178} 416-422}.
\\
Barnes, T., Swanson, E.S. (1992).
     {\it Phys. Rev.} {\underline {D 46}}, 131-159.
\\
Kharzeev, D. and Satz, H.(1994). 
      {\it Phys. Lett.} {\underline {B 334}}, 155-162.
\\
Martins, K., Blaschke, D., Quack, E. (1995).
        {\it Phys. Rev.} {\underline {C 51}}, 2723-2738.
\\
Wong, C.-Y. (1995).
    {\it preprint ORNL-CTP-95-04, hep-ph/9506270}.
\\
Alde, D.M. {\it et al.}, (E772 Collaboration) (1991).
 {\it Phys. Rev. Lett.} {\underline 66}, 133-136.
\\
Ramos, S. {\it et al.}, (NA38 Collaboration) (1995).
{\it Nucl. Phys.} \underline{A590}, 117c-126c.

\newpage
\begin{figure}[b]
\centerline{
\begin{minipage}[b]{15cm}
\centerline{
\psdraft   
\psfig{figure=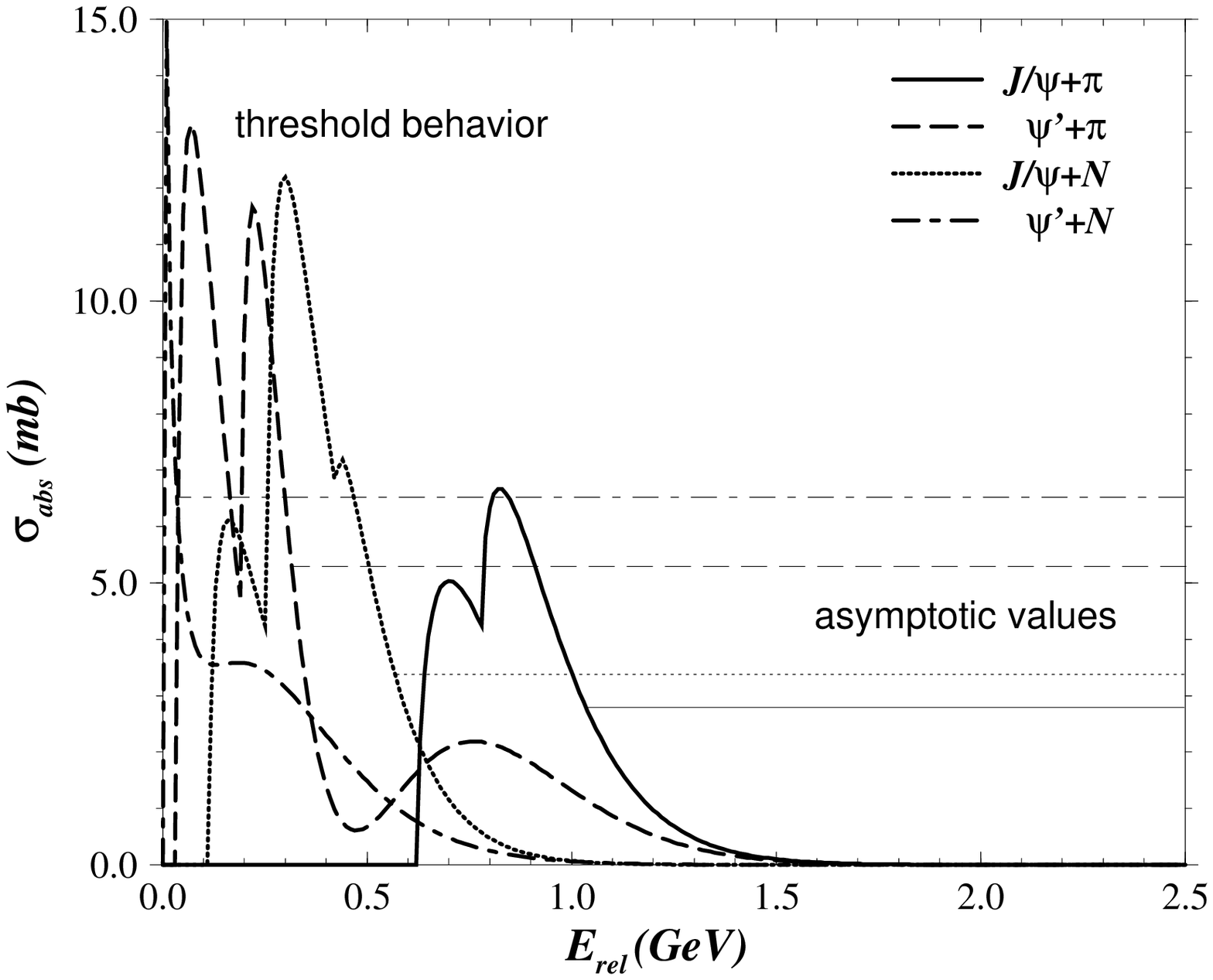,width=15cm,height=12cm}
}
\vspace{-1.3cm}
\parbox{0.cm}{} 
{        \renewcommand{\baselinestretch}{1.}
\footnotesize    
\centerline{
Fig.\,1. 
\begin{minipage}[t]{10cm}
        {
        Cross sections for the absorption of $J/\psi$ and $\psi'$ mesons
        by pions and nucleons near the reaction threshold and at high 
        energies.
        }
      \end{minipage}
}
\vspace{1.cm}
}
\end{minipage}
}
\end{figure}
\newpage
\begin{figure}[b]
\centerline{
\begin{minipage}[b]{15cm}
\vspace{0.5cm}
\centerline{
\psdraft   
    \psfig{figure=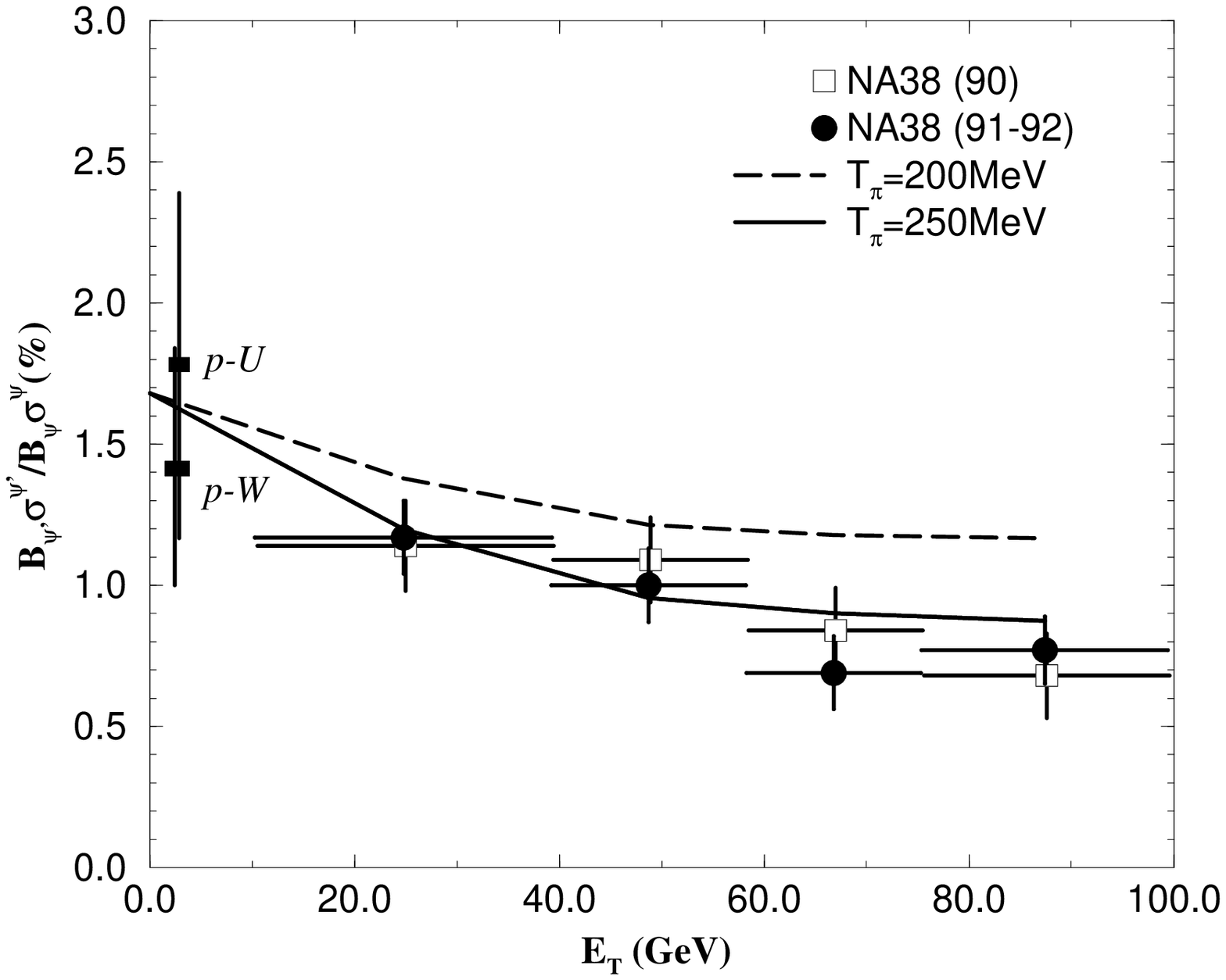,width=15cm,height=11cm}
}
\vspace{-1.3cm}
\parbox{0.cm}{}
{       \renewcommand{\baselinestretch}{1.}
\footnotesize
\centerline{
Fig.\,2.
\begin{minipage}[t]{10cm}
   $\psi'/\psi$ ratio in S-U collisions at 200 GeV A. Curves are shown for 
    two pion gas temperatures.
      \end{minipage}
}
}
\end{minipage}
}
\end{figure}

\end{document}